\documentclass[11pt,twoside]{article}
\usepackage{asp2010}
\usepackage{color}

\resetcounters

\begin{document}

\title{\LARGE A public database for white dwarf asteroseismology
with fully evolutionary models} 

\vskip -4mm
\title{I. Chemical profiles and pulsation periods of ZZ Ceti (DAV) 
stars}

\author{Alejandra D. Romero$^{1,2}$,  Alejandro H. C\'orsico$^{1,2}$, 
Leandro G. Althaus$^{1,2}$, \& Marcelo M. Miller Bertolami$^{1,2}$}
 
\affil{$^{1}$Facultad de Ciencias Astron\'omicas y Geof\'isicas, 
           Universidad Nacional de La Plata, 
           Paseo del Bosque s/n, 
           (1900) La Plata, 
           Argentina}
\affil{$^{2}$Consejo Nacional de Investigaciones Cient\'ificas y T\'ecnicas 
(CONICET)}

\vskip 4mm
{\small Emails: \url{aromero, acorsico, althaus, mmiller@fcaglp.unlp.edu.ar}}

\begin{abstract}
We present  a large  bank of chemical  profiles and  pulsation periods
suited for  asteroseismological studies of  ZZ Ceti (or  DAV) variable
stars.   Our background  equilibrium  DA white  dwarf  models are  the
result of  fully evolutionary computations that take  into account the
complete history of the progenitor stars from the ZAMS. The models are
characterized by self-consistent chemical structures from the centre to
the  surface, and  cover a  wide  range of  stellar masses,  effective
temperatures,  and  H envelope  thicknesses.   We  present dipole  and
quadrupole  pulsation   $g$-mode  periods  comfortably   covering  the
interval of periods observed in ZZ Ceti stars.

Complete tabulations of chemical  profiles and pulsation periods to be
used in  asteroseismological period fits, as well  as other quantities
of  interest,  can  be  freely  downloaded  from   our  website 
(\url{http://www.fcaglp.unlp.edu.ar/evolgroup}).
\end{abstract}

\section{Introduction}

White  dwarf asteroseismology  is a  powerful astrophysical  tool that
fully exploits  the comparison between the  observed pulsation periods
in white  dwarfs and the periods computed  for appropriate theoretical
models.   It  allows us  to  infer  details  of the  origin,  internal
structure  and  evolution of  white  dwarfs  (Winget  \& Kepler  2008;
Althaus et  al.  2010a).  ZZ Ceti (or  DAV) stars constitute  the most
numerous  group  of degenerate  variable  stars.   They are  otherwise
normal   DA   (H-rich   atmospheres)   white   dwarfs   that   exhibit
$g$(gravity)-mode  pulsations.  Recently,  our  group 
\textcolor{red}{\bf La  Plata Stellar Evolution  and Pulsation Research Group}
has performed for
the first time a detailed  asteroseismological study on an ensemble of
44 bright ZZ Ceti stars  by employing fully evolutionary (that is, non
static) DA white  dwarf models (Romero et al.  2012, Romero 2012). Our
asteroseismological approach basically consists in the employment of a
large suite of detailed  stellar models characterized by very accurate
and  updated  physical  ingredients.   These models  were  produced  by
computing  the complete evolution  of the  progenitor stars.   We have
applied successfully this approach to the hot GW Vir (or DOV) stars in
the past (see  C\'orsico et al. 2007a, 2007b,  2008, 2009).  Since the
final chemical stratification of white dwarfs is fixed in prior stages
of their evolution, the evolutionary history of progenitor stars is of
utmost importance in the context of white dwarf asteroseismology.

Here, we  present a  complete set of  pulsational results that  can be
useful to  perform asteroseismological  studies of ZZ Ceti  stars.  These
include   the  internal  chemical   profiles  and   the  run   of  the
Brunt-V\"ais\"al\"a and  Lamb frequencies, as  well as a large  set of
adiabatic   pulsation  periods,   time-averaged   oscillation  kinetic
energies,  and  first order  rotational  splitting coefficients.   All
these quantities can be freely  downloaded from our website. 

This database can be used in two ways: (1) carrying out period fits 
using directly the bank of periods we provide, corresponding to our 
set of equilibrium DA white dwarf models, or alternatively, (2) 
by scaling our internal
chemical profiles to the structure of other independent DA white dwarf 
models, and then by performing period fits using
the pulsation periods computed for such models. 

Below, we
summarily describe  the input physics of our models and  the pulsation 
computations, and then we present the format of our database.

\section{Input physics and evolutionary computations}

The state-of-the-art DA white dwarfs evolutionary models employed have
been computed with the LPCODE  evolutionary code. Details of this code
and the input physics that this code includes can be found in Althaus et
al. (2010b), Renedo  et al. (2010) and references  therein.  Below, we
briefly enumerate the main physical ingredients:

\begin{itemize}

\item For the high-density  regime characteristics of white dwarfs,
  we have used the equation of state (EoS) of Segretain et al. (1994).

\item For the low-density regime, we used an updated version of the
  EoS of Magni \& Mazzitelli (1979).

\item Radiative  opacities are from  the OPAL project  (Iglesias \&
  Rogers  1996),   including  carbon-  and   oxygen-rich  composition,
  supplemented  at low  temperatures with  the molecular  opacities of
  Alexander \& Ferguson (1994).

\item Conductive opacities are those of Cassisi et al. (2007).

\item Neutrino emission rates are taken from Itoh et al. (1996) and
  Haft et al. (1994).

\item  The  $^{12}$C($\alpha,   \gamma)^{16}$O  reaction  rate,  of
  special relevance for the C-O  stratification of the white dwarf, is
  that of Angulo et al. (1999).

\item  Convection  has  been  modeled  with the  formalism  of  the
  mixing-length theory as given by the ML2 parametrization (Tassoul et
  al. 1990).

\item Treatment of the chemical profiles. We  have   considered  the   
distinct  physical  processes   that  are responsible for changes in  
the chemical abundance distribution during
white  dwarf  evolution:

\begin{itemize}

\item[--]  Element   diffusion:  it  strongly   modifies  the  chemical
  composition  profile   throughout  their  outer   layers.   We  have
  considered gravitational  settling as  well as thermal  and chemical
  diffusion  of $^1$H, $^3$He,  $^4$He, $^{12}$C,  $^{13}$C, $^{14}$N,
  and $^{16}$O.  The treatment of time-dependent diffusion is based on
  the    multicomponent   gas    treatment   presented    in   Burgers
  (1969). Diffusion  becomes operative once the wind  limit is reached
  at high effective  temperatures (Unglaub \& Bues 2000).  As shown in
  previous studies,  diffusion in  the white dwarf  envelope is  a key
  ingredient as far  as the mode-trapping properties of  ZZ Ceti stars
  is concerned.

\item[--]  Abundance changes  resulting from  residual  nuclear burning
  (mostly during  the hot stages  of white dwarf evolution)  have been
  taken into account.  Nuclear burning fixes the maximum value  
  of $M_{\rm H}$ that is expected in a DA white dwarf.

\item[--] Chemical re-homogenization  of the inner carbon-oxygen profile
  induced  by   Rayleigh-Taylor  instabilities  has   been  considered
  following Salaris et al.  (1997).  These instabilities arise because
  of  the positive molecular  weight gradients  that remain  above the
  flat chemical profile left by convection during core helium burning.

\end{itemize}
\end{itemize}

\section{Pulsation computations}

\begin{figure*} 
\begin{center}
\includegraphics[clip,width=12 cm]{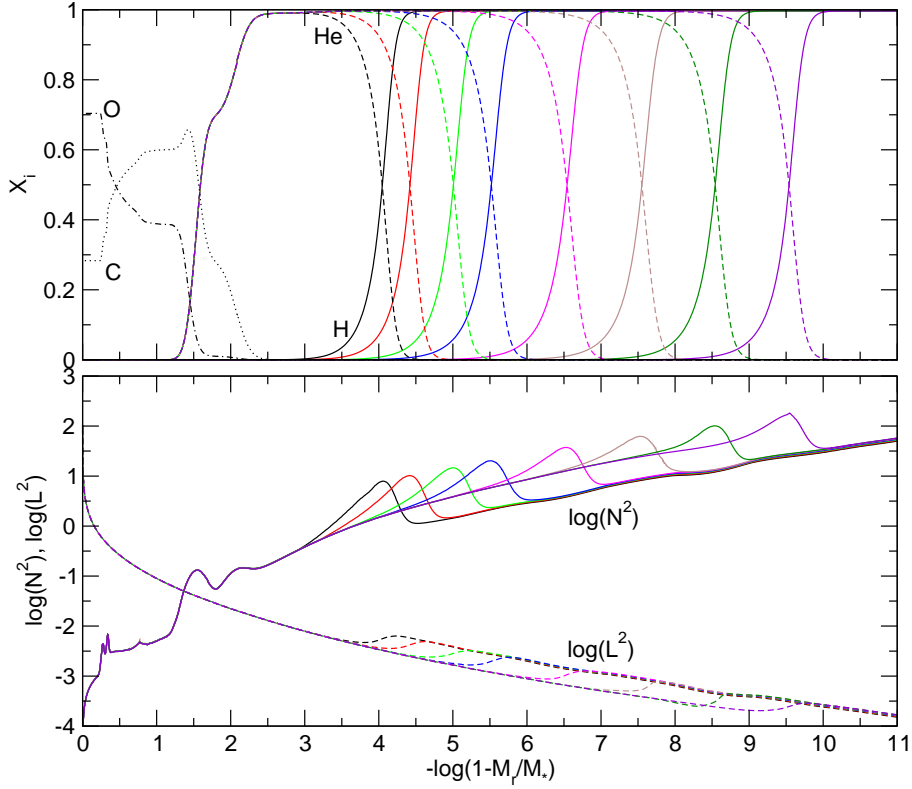} 
\caption{Upper panel:  example of  the internal chemical  profiles for
  $^1$H, $^4$He, $^{12}$C, and $^{16}$O  of our DA white dwarf models.
  The cases  shown in the figure  correspond to models  with a stellar
  mass of $M_*= 0.593  M_{\odot}$, an effective temperature of $T_{\rm
    eff}\sim 12\,350$  K, and  H envelope thicknesses  of $\log(M_{\rm
    H}/M_*)= -3.93$  (black), $-4.28$ (red),  $-4.85$ (green), $-5.34$
  (blue),  $-6.33$ (magenta), $-7.34$  (brown), $-8.33$  (dark green),
  and  $-9.33$ (violet).  Lower  panel: the  logarithm of  the squared
  Brunt-V\"ais\"al\"a and Lamb  frequencies corresponding to the cases
  depicted  in  the  upper  panel.  The  correspondence  between  each
  bump-like feature  in $N^2$ and  the chemical transition  regions of
  the models is clearly visible.}
\label{figure1} 
\end{center}
\end{figure*}

We have employed the pulsation  code described in C\'orsico \& Althaus
(2006),  which  is  coupled  to  the LPCODE  evolutionary  code.   The
pulsation  code is based  on a  general Newton-Raphson  technique that
solves  the  full  fourth-order  set of  equations  governing  linear,
adiabatic,  nonradial stellar  pulsations following  the dimensionless
formulation  of  Dziembowski (1971)  (see  Unno  et  al.  1989).   The
pulsation  code provides  the dimensionless  eigenfrequency $\omega_k$
($k$  being the  radial order  of the  mode) and  eigenfunctions $y_1,
\cdots,  y_4$.   From  these   basic  quantities,  the  code  computes
pulsation   periods  ($\Pi_k$),   time-averaged   oscillation  kinetic
energies  ($E_k$),  rotation  splitting coefficients  ($C_k)$,  weight
functions  ($W_k$),  and  variational  periods  ($\Pi^{\rm v}_k$)  for  each
computed eigenmode (see Appendix A of C\'orsico \& Althaus 2006 for the
definition  of  these quantities).  Usually,  the relative  difference
between $\Pi^{\rm v}_k$  and $\Pi_k$ is lower than  $\approx 10^{-4}$, which
represents the typical uncertainties  in our theoretical periods.  The
prescription we  follow to assess  the run of  the Brunt-V\"ais\"al\"a
frequency  ($N$) for  a  degenerate environment  typical  of the  deep
interior  of  a  white  dwarf  is the  so-called  ``Ledoux  Modified''
treatment (Tassoul et al.  1990), appropriately generalized to include
the  effects of  having  three nuclear  species  (oxygen, carbon,  and
helium) varying in  abundance in the same place of  the star (a triple
chemical interface).   Fig. \ref{figure1}  displays an example  of the
chemical  profiles  of  our  models  with a  fixed  stellar  mass  and
effective temperature and for  different thicknesses of the H envelope
(upper panel),  and the run  of the Brunt-V\"ais\"al\"a and  Lamb (for
$\ell= 1$) frequencies (lower panel).   Details can be found in Romero
et al.  (2012) and Romero (2012).

\begin{figure*} 
\begin{center}
\includegraphics[clip,width=12 cm]{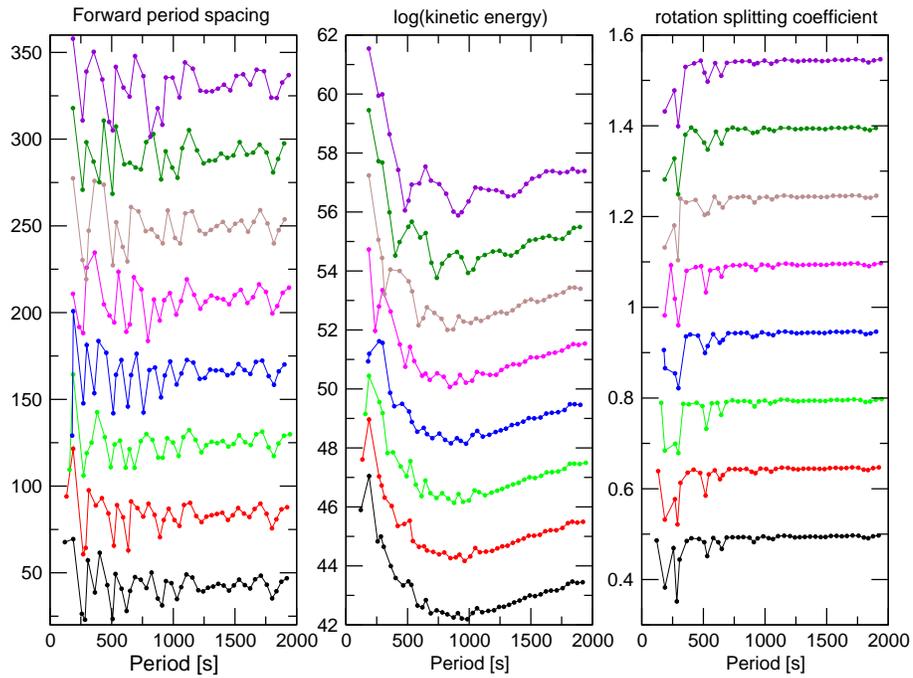} 
\caption{Pulsation quantities for $\ell= 1$ modes corresponding to the same 
models analyzed in Fig. \ref{figure1}. In each panel, 
the curves have been arbitrarily shifted upwards for clarity,
except the lowest one. Left panel: forward period 
spacing ($\Delta \Pi_k= \Pi_{k+1}-\Pi_k$) vs periods; middle panel: 
oscillation kinetic energy ($\log(E_k)$) vs periods; right panel: 
first order rotation splitting coefficient ($C_k$) vs periods.} 
\label{figure2} 
\end{center}
\end{figure*}

We  present results  for DA  white dwarf  models with  stellar masses,
effective temperatures and H envelope thicknesses in the ranges $0.525
\leq M_*/M_{\odot}  \leq 0.878$, $9000 \leq T_{\rm  eff} \leq 14\,000$
K, and $-3.62 \leq  \log(M_{\rm H}/M_*) \leq -9.34$, respectively.  We
have performed  pulsation calculations on  about ($11 \times  7 \times
200)  = 15\,400$  DA  white dwarf  models.  In this  account, we  have
considered the  number of  stellar mass values  ($11$), the  number of
thicknesses of the H envelope for each sequence ($\approx 7$), and the
number  of models ($\approx  200$) with  effective temperature  in the
interval $14\, 000 - 9000$  K, respectively. For each model, adiabatic
pulsation $g$-modes with  $\ell = 1$ and $2$ and  periods in the range
$80-2000$ s have been computed.  This range of periods corresponds (on
average) to $1 \lesssim k \lesssim 50$ for $\ell= 1$ and $1 \lesssim k
\lesssim  90$  for $\ell=  2$.  So, more  than  $\sim  2 \times  10^6$
adiabatic  pulsation periods  have been  computed. An  example  of our
pulsation results is depicted in Fig. \ref{figure2}, where we show the
forward  period  spacing,  the  oscillation kinetic  energy,  and  
the first order rotational splitting coefficients   for   the    
same   models    analyzed   in Fig.  \ref{figure1}. Note  the  gradual 
changes  experienced by  these quantities as we vary the value of the H 
envelope thickness.

In Table \ref{table1} we show the grid of evolutionary sequences 
included in our database. The values of the stellar mass of our set 
of DA white dwarf models are shown in the upper row. 
The masses of H corresponding to the different envelope thicknesses 
for each stellar mass are shown in rows 2 to 9. Row 2 corresponds 
to the maximum value of the thickness of the H envelope
for each stellar mass according to our evolutionary computations 
(``canonical envelopes''). Rows 3 to 9 correspond to H envelopes
thinner than the canonical ones.

\begin{table*}
\caption{}
\centering
\scalebox{0.9}[0.9]{
\hspace{-7mm}
\begin{tabular}{rccccccccccc}
\hline
\hline
$M_*/M_{\odot}$ &  0.5249 &0.5480 &0.5701 &0.5932 &0.6096 &0.6323 &0.6598 &0.7051&0.7670 & 0.8373 &0.8779 \\
\hline
$\log\left(\frac{M_{\rm H}}{M_*}\right)$ & -3.62  & -3.74  & -3.82  & -3.93  & -4.02  & -4.12  & -4.25 & -4.45  & -4.70  & -5.00  & -5.07 \\
\hline
                                        & -4.27  & -4.27  & -4.28  & -4.28  & -4.45  & -4.46  & -4.59  &       &   &        &       \\
                                        & -4.85  & -4.85  & -4.84  & -4.85  & -4.85  & -4.86  & -4.87  & -4.88 & -4.91  &        &       \\
                                        & -5.35  & -5.35  & -5.34  & -5.34  & -5.35  & -5.35  & -5.35  & -5.36  & -5.37  & -5.41  & -5.40 \\
                                        & -6.33  & -6.35  & -6.33  & -6.33  & -6.34  & -6.34  & -6.35  & -6.35  & -6.35  & -6.36  & -6.39 \\
                                        & -7.34  & -7.33  & -7.34  & -7.34  & -7.33  & -7.35  & -7.33  & -7.35  & -7.34  & -7.36  & -7.38 \\
                                        & -8.33  & -8.33  & -8.31  & -8.33  & -8.33  & -8.33  & -8.33  & -8.34  & -8.33  & -8.34  & -8.37 \\
                                        & -9.25  & -9.22  & -9.33  & -9.33  & -9.25  & -9.34  & -9.33  & -9.34  & -9.33  & -9.34  & -9.29 \\
\hline 
\label{table1}
\end{tabular}
}
\end{table*}

\section{Format of the files to be downloaded}

The files of our database can be downloaded from our website: 

\vskip 3mm
\noindent \url{http://www.fcaglp.unlp.edu.ar/evolgroup/TRACKS/PULSATIONS/PULSATIONS_DA/pulsations_cocore.html}

\vskip 3mm
\noindent Once in  the site, you can see that the  
files are organized in  three separate tables,  which  have  exactly  
the  same form  as  Table  \ref{table1}
above. For  a given  table, each element  is associated to  a specific
sequence  characterized  by  the  corresponding values  of  $M_*$  and
$\log(M_{\rm  H}/M_*)$.  In the  first  table  we  provide results  of
periods   ($\Pi_k$)  and   the  other   quantities   ($\Delta  \Pi_k$,
$\log(E_k)$ and $C_k$) for the  harmonic degree $\ell= 1$, in terms of
the radial order  $k$.  The same information, but  for the case $\ell=
2$,  is  provided in  the  second  table.   Finally, the  third  table
includes   the   squared    Brunt-V\"ais\"al\"a   ($N^2$)   and   Lamb
($L_{\ell}^2$) frequencies,  the Ledoux  term ($B$), and  the chemical
abundances (by mass) of hydrogen ($X_{\rm H}$), helium ($X_{\rm He}$),
carbon ($X_{\rm C}$), and oxygen  ($X_{\rm O}$) in terms of the radial
coordinate   ($r/R_*$)  and  the   outer  mass   fraction  coordinate,
[$-\log(1-M_r/M_*)$].

\begin{figure*} 
\begin{center}
\includegraphics[clip,width=14 cm]{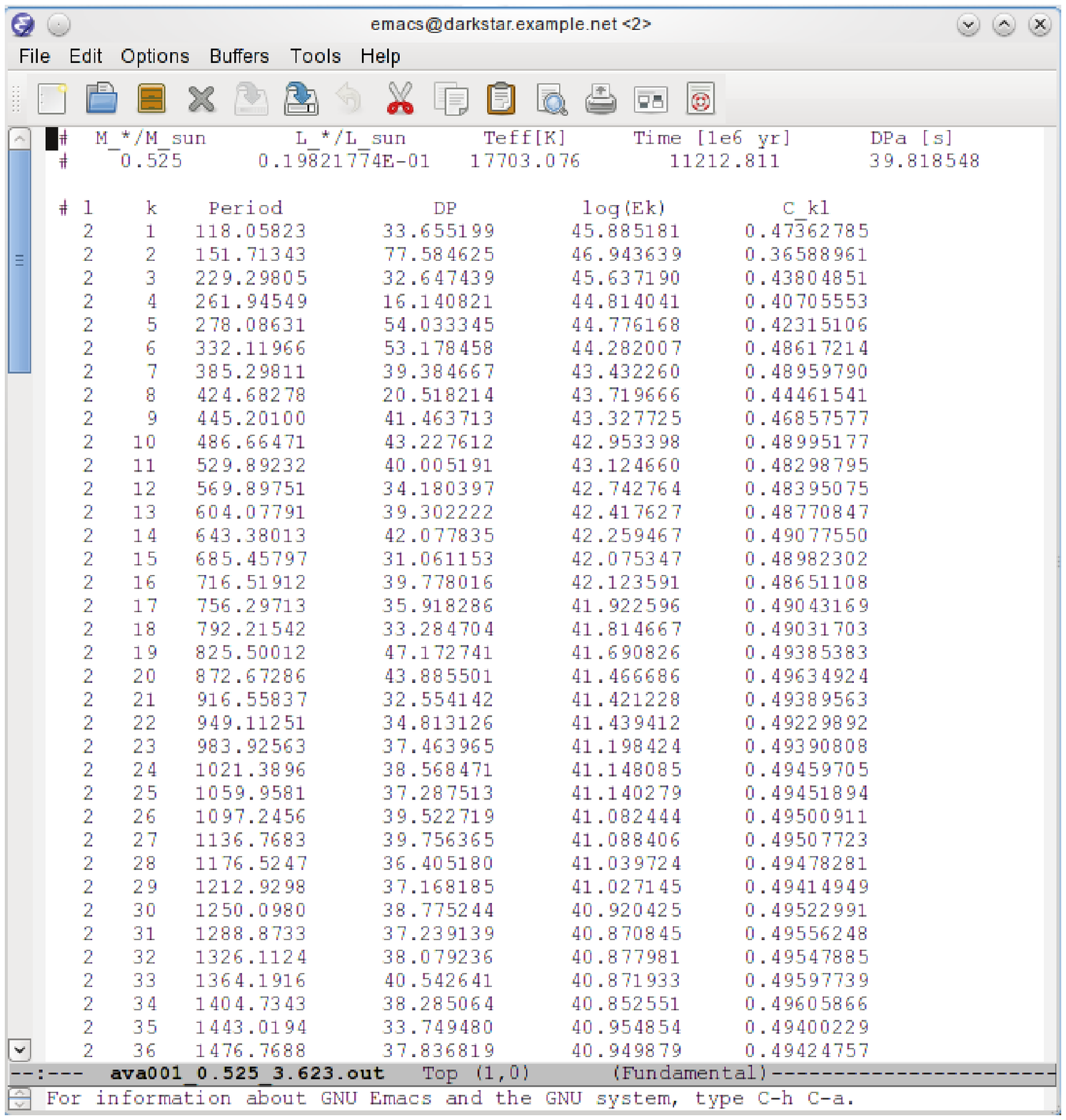} 
\caption{The       aspect       of       the      file       {\protect
    \url{ava001_0.525_3.623.out}}, corresponding  to a DA  white dwarf
  model  with $M_*=  0.525 M_{\odot}$,  $L_*/L_{\odot}=  1.9822 \times
  10^{-2}$,  $T_{\rm  eff}= 17\,  703$  K,  $\tau= 11212.811 \times 10^{6}$ yr, 
and $\log(M_{\rm  H}/M_*)=   -3.623$.}
\label{figure3} 
\end{center}
\end{figure*} 

\subsection{The first table: periods and other quantities for $\ell= 1$}

Each element of the table  is a hyperlink that leads to a given
tar  gzipped file. You  can download  a specific  tar gzipped  file by
simply  clicking on a  given element  of the  table.  For
instance, if you  click at element \textcolor{blue}{\underline{\bf -3.62}}  
in the first column of the table,  you will download  the 
file:  \url{ava_0.525_3.623.tgz}.  This
file  corresponds to  the  sequence with  $M_*=  0.525 M_{\odot}$  and
$\log(M_{\rm H}/M_*)=  -3.623$. After you  detar the file,  you should
obtain a sequence of files:

$$
\begin{array}{l}
\mbox{\url{ava001_0.525_3.623.out}}\\
\mbox{\url{ava002_0.525_3.623.out}}\\
\mbox{\url{ava003_0.525_3.623.out}}\\
\mbox{\url{ava004_0.525_3.623.out}}\\
\vdots\\
\end{array}
$$

\noindent Each  of these files corresponds to  a different, decreasing
effective temperature. For instance, in Fig. \ref{figure3} we show the
appearance  of  the  file \url{ava001_0.525_3.623.out}.   The  heading
contains   the   value   of   the   stellar  mass   in   solar   units
(\url{M_*/M_sun}),   the    stellar   luminosity   in    solar   units
(\url{L_*/L_sun}),   the  effective   temperature  in   Kelvin  (\url{
  Teff[K]}), the age in $10^6$  yr units (\url{Time[1e6 yr]}), and the
asymptotic period  spacing in seconds (\url{DPa [s]}),  computed as in
Tassoul et  al. (1990).  Below  the heading, the  following quantities
are listed: the harmonic degree (\url{l}), the radial order (\url{k}),
the  periods (\url{Period}), the  forward period  spacings (\url{DP}),
the logarithm of the oscillation kinetic energies (\url{log(Ek)}), and
the first order rotation splitting coefficients (\url{C_kl}).

A very important quantity such  as the rate of period change of
a  given pulsation  mode with  radial order  $k$  ($\dot{\Pi}_k \equiv
d\Pi_k/dt$),  can be  easily computed  from the  age ($\tau$)  and the
period ($\Pi_k$) of  a given model ($j$) and  the corresponding values
of the previous model ($j-1$):

$$
\frac{d\Pi_k}{dt}= \frac{\Pi_k^j-\Pi_k^{j-1}}
{\tau^j-\tau^{j-1}}
$$

Finally, we  also give  the option  of downloading  the {\sl  complete}  set of
periods   (for  all   stellar  masses,   H  envelopes   and  effective
temperatures) at once in the hyperlink below the table.

\subsection{The second table: periods and other quantities for $\ell= 2$}

All the above explanation holds also for the case of the second table,
but in this case, the results correspond to $\ell= 2$ pulsation modes.

\subsection{The third table: critical frequencies and chemical profiles}

Now, if you  click (for instance) at element  
\textcolor{blue}{\underline{\bf -3.62}} in
the    first column of the  table,    you     will    download    the     file:
\url{par_0.525_3.623.tgz}.   Again,  this   file  corresponds  to  the
sequence  with  $M_*=   0.525  M_{\odot}$  and  $\log(M_{\rm  H}/M_*)=
-3.623$. After  you detar  the file, you  should obtain a  sequence of
files:

$$
\begin{array}{l}
\mbox{\url{par001_0.525_3.623.out}}\\
\mbox{\url{par002_0.525_3.623.out}}\\
\mbox{\url{par003_0.525_3.623.out}}\\
\mbox{\url{par004_0.525_3.623.out}}\\
\vdots\\
\end{array}
$$

\begin{figure*} 
\begin{center}
\includegraphics[clip,width=15 cm]{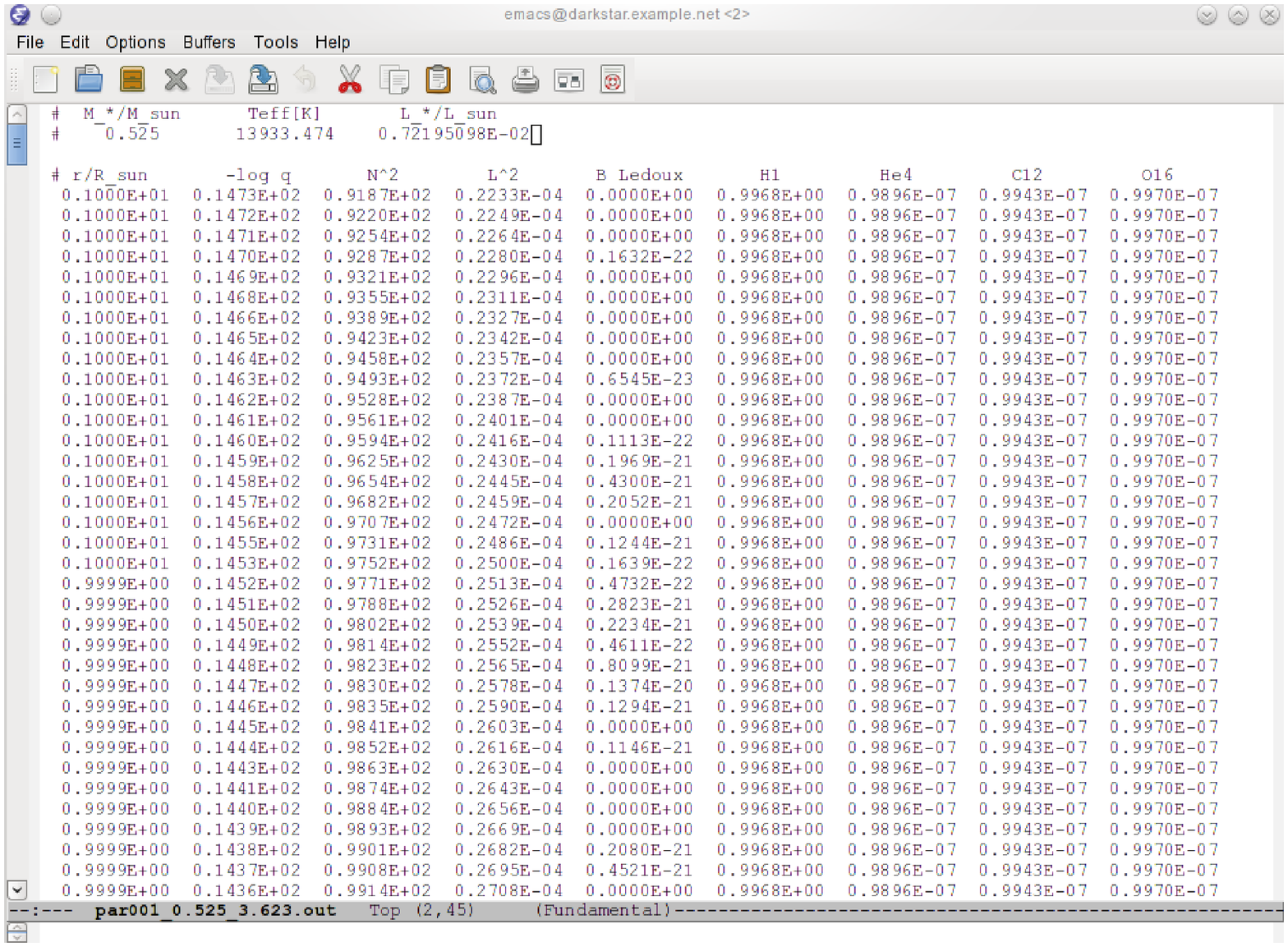} 
\caption{The       aspect       of       the      file       {\protect
    \url{par001_0.525_3.623.out}}, corresponding  to a DA  white dwarf
  model  with  $M_*=  0.525  M_{\odot}$, $L_*/L_{\odot}=  7.22  \times
  10^{-3}$,  $T_{\rm  eff}= 13\,  933$  K, and $\log(M_{\rm  H}/M_*)=   -3.623$.}
\label{figure4} 
\end{center}
\end{figure*}

\noindent  Again,  each of  these  files  correspond  to a  different,
decreasing effective  temperature. But, at  variance with the  cases of
the  first and second  tables, in  this case  the interval  in $T_{\rm
  eff}$   between  consecutive   files  is   quite  large
($\sim 500$ K). In  Fig.
\ref{figure4}    we    show     the    appearance    of    the    file
\url{par001_0.525_3.623.out}.  The  parameters at the  heading are, in
this case, the stellar mass, the stellar luminosity, and the effective
temperature.  Below the heading,  the following quantities are listed:
the  normalized  radial coordinate,  (\url{r/R_sun}),  the outer  mass
fraction  coordinate (\url{-log  q}), the  squared Brunt-V\"ais\"al\"a
frequency  (\url{N^2}), the  squared Lamb  frequency  (\url{L^2}), the
Ledoux term  (\url{B Ledoux}),  and finally the  abundance by  mass of
hydrogen  (\url{H1}),  helium  (\url{He4}),  carbon  (\url{C12}),  and
oxygen (\url{O16}).

\section{Final remarks}

\begin{itemize}

\item{}  Additional results for  other values  ​​of  the H
  envelope thickness  that are not  included in this database,  can be
  obtained by request to the authors.

\item{} If  you use this  database and publishes your  results, please
  cite the  paper: {\it ``Toward ensemble asteroseismology  of ZZ Ceti
    stars with fully evolutionary models''}, Romero, A. D., C\'orsico,
  A. H.,  Althaus, L.  G., Kepler,  S. O., Castanheira,  B. G., Miller
  Bertolami, M. M. 2012, MNRAS, 420, 1462

\item Any  comment/criticism that helps to improve  this database will
  be greatly appreciated!

\end{itemize}

{}

\end{document}